\documentclass[manuscript,screen]{acmart}
\usepackage[capitalize]{cleveref}

\usepackage{natbib}
\usepackage{adjustbox}
\makeatletter
\newcommand\textlist[4]{%
  \let\last@item\relax
  \let\last@sep\relax
  \@for\@ii:=#4\do{%
    \ifx\last@item\relax\else
      \ifx\last@sep\relax
        \def\last@sep{#2}%
      \else#1\fi
      #3{\last@item}%
    \fi
    \let\last@item\@ii
  }%
  \ifx\last@item\relax\else
    \last@sep#3{\last@item}%
  \fi
}
\makeatother

\usepackage{xcolor}
\newcommand{\juba}[1]{}
\newcommand{\chinasa}[1]{}
\newcommand{\michelle}[1]{}

\usepackage{xcolor}
\renewcommand{\juba}[1]{\textcolor{red}{[Juba: #1]}}
\renewcommand{\chinasa}[1]{\textcolor{blue}{[Chinasa: #1]}}
\renewcommand{\michelle}[1]{\textcolor{orange}{[Michelle: #1]}}

\AtBeginDocument{%
  \providecommand\BibTeX{{%
    \normalfont B\kern-0.5em{\scshape i\kern-0.25em b}\kern-0.8em\TeX}}}

\setcopyright{acmcopyright}
\copyrightyear{2023}
\acmYear{2023}
\acmConference[EAAMO'23]{Conference on Equity and Access in Algorithms, Mechanisms, and Optimization}
\begin{document}

\title[Enabling Patient Agency in the Use of AI-Enabled Healthcare]{IAC: A Framework for Enabling Patient Agency in the Use of AI-Enabled Healthcare}

\author{Chinasa T. Okolo}
\authornote{Both authors contributed equally to this research.}
\email{chinasa@cs.cornell.edu}
\orcid{0000-0002-6474-3378}
\affiliation{%
  \institution{Cornell University}
  \streetaddress{350 Gates Hall}
  \city{Ithaca}
  \state{New York}
  \postcode{14853}
  \country{United States}
}

\author{Michelle González Amador}
\authornotemark[1]
\email{mgonzalez@merit.unu.edu}
\orcid{0000-0003-0265-4545}
\affiliation{%
  \institution{UNU-MERIT and Maastricht University}
  \streetaddress{24 Boschstraat}
  \city{Maastricht}
  \postcode{6211 AX}
  \country{Netherlands}
}

\renewcommand{\shortauthors}{Okolo and González Amador}

\begin{abstract}
  In healthcare, the role of AI is continually evolving, and understanding the challenges its introduction poses on relationships between healthcare providers and patients will require a regulatory and behavioral approach that can provide a guiding base for all users involved. In this paper, we present IAC (Informing, Assessment, and Consent), a framework for evaluating patient response to the introduction of AI-enabled digital technologies in healthcare settings. We justify the need for IAC with a general introduction of the challenges with and perceived relevance of AI in human-welfare-centered fields, with an emphasis on the provision of healthcare. The framework is composed of three core principles that guide how healthcare practitioners can inform patients about the use of AI in their healthcare, how practitioners can assess patients' acceptability and comfortability with the use of AI, and how patient consent can be gained after this process. We propose that the principles composing this framework can be translated into guidelines that improve practitioner-patient relationships and, concurrently, patient agency regarding the use of AI in healthcare while broadening the discourse on this topic.
\end{abstract}




\maketitle

\section{Introduction}
One of the defining characteristics of modern societies is the embeddedness of technology in many aspects of human life. Understanding how humans respond to technological innovations is essential for the successful introduction of digital tools in human-welfare-centered fields such as healthcare, education, and labor markets. In healthcare, the role of artificial intelligence (AI) is continually evolving, and understanding the paradigm shift of traditional medical relationships from physician-patient to physician-AI-patient will require a comprehensive awareness of existing human-human interactive structures in medicine and the challenges the introduction of AI poses to them. We find that a technological approach to healthcare comes with its own set of challenges. In particular, we focus on the complex social structure that emerges from physician-AI-patient interaction. Studies that look at users’ responses to the introduction of AI-enabled tools in human-welfare-centered fields indicate the outcome of the interaction is contingent on the level of complexity and type of the task, the expertise of the user, and the mental models users have formed about the fairness of the tool \cite{kaufmann2020teachers,lee2018understanding,sundar2001conceptualizing,cowgill2018bias}. In healthcare, past experiences with technology and AI complexity mediate patient acceptance of AI \cite{richardson2022framework, richardson2021patient}. Begetting mental models that ignite a cooperative interaction between healthcare providers and patients will require a regulatory and behavioral approach that can provide a guiding base for all users involved \cite{emanuel2019artificial,terry2019regulating}.

We expand work exploring the ethical concerns of AI in healthcare, patient perception of AI tools, and the impact of trust and privacy in technology acceptance for healthcare by developing IAC, a framework to guide the introduction of AI-enabled technologies in healthcare settings and shift how patients leverage their agency to consent to the use of these tools in their healthcare. Our framework aims to standardize the practice of evaluating patient reception to AI integration within healthcare and establish the foundation for regulatory policies that will enforce the fair use of AI systems in accordance with existing healthcare standards such as the Health Insurance Portability and Accountability Act of 1996 (HIPAA) in the United States and the UK's Data Protection Act 2018. This paper follows a human-welfare or patient-centered approach as prescribed by \citet{chen2022acceptance, musbahi2021public, richardson2021patient, richardson2022framework, young2021patient} for the implementation of AI usage guidelines and ethical protocols, and develops a unique framework with a strong behavioral base. An important contribution of IAC is its focus on the socialization of standardized ethical procedures, which, to our knowledge, has not yet been explored in the literature. 
\section{Related Work}

\subsection{AI in healthcare}
In the provision of healthcare services, artificial intelligence (AI) is ubiquitous. With applications in medical imaging \cite{zhou2019artificial, kwan2000automatic, roblot2019artificial, bhanumurthy2014automated}, personalized medicine~\cite{awwalu2015artificial, schork2019artificial}, drug discovery~\cite{altae2017low, jing2018deep, klucznik2018efficient, feinberg2018potentialnet, wu2019admet}, epidemiology \cite{barrett2008interaction, apolloni2009computational, wakamiya2018twitter}, and operational efficiency \cite{rigby2019ethical}, AI is undeniably changing the healthcare landscape for the better. Within this domain, literature discussing the ethical concerns, perceptions of, and implications of AI in healthcare has become more prominent. For example, \citet{mccradden2020ethical} analyze perspectives from adult patients with brain cancer regarding issues of consent around their health data, the allocation of health resources through computational methods, and privacy concerns associated with using their data for health research  \citet{liyanage2019artificial} survey healthcare informaticians and clinicians on the benefits, risks, potential of adoption, implications, and the future of AI in primary care. \citet{lai2020perceptions} interview a range of French professionals with an interest in or experience with AI for healthcare (physicians, AI researchers, AI consultants, ethicists, public health researchers, etc.) to gain insight into the needs, opportunities, and challenges arising from the use of AI in healthcare. Further work done by \citet{xiang2020implementation} examines the difference in AI sentiment between healthcare and non-healthcare workers regarding the safety of AI, consistency between physician and AI-based diagnosis, receptivity to AI being used in their care, the respective demand for AI in healthcare, and opportunities for AI to improve operational aspects of patient care. This corpus shows that as AI becomes more persistent throughout healthcare, it is equally as important to study the social and ethical impact AI has within healthcare along with the technical specifications of AI-enabled technologies.

\subsection{Clinician-patient interaction with AI}
The debate surrounding the effect of technological innovations on the clinician-patient relationship is not new \cite{boucher2010technology, botrugno2019information}. Studies show a mixed perspective, with most research agreeing on the management, coordination, and efficiency benefits of technology in healthcare \cite{mandl1998electronic, boucher2010technology, gustafson2001effect}, while the strand of research interested in the interaction between physicians and patients documents the challenges that emerge with digital means of healthcare communication\cite{mandl1998electronic} and healthcare provision \cite{murray2003impact}. For instance, \citet{botrugno2019information} discusses that while Information Technologies (IT) may ease communication between a clinician and her patient, it could also unintentionally \textit{dehumanize} it and worsen it since the lack of physical presence decreases the number of cues a doctor has to evaluate her patient's health. 

In the same vein, concerns arise about service providers’ ability to engage in positive cooperation between clinicians, AI, and patients \cite{taddeo2018ai,terry2016emerging,rigby2019ethical,aminololama2019doctor}. \citet{terry2016emerging} put forth the idea that the increasing use of digital means of healthcare provision is eroding clinician-patient empathy and the associated positive health outcomes that come from this interaction. Moreover, \citet{morley2020ethics} warn that an over-reliance or mismanagement of AI can lead to the impersonalization of healthcare provision, and consequently a decrease in trust and clinician-patient empathy \cite{morley2020ethics, juengst2016personalized}. This is compounded by the fact that the line that separates AI tools in healthcare as ``decision support systems'' and ``decision-making systems'' can be blurry under certain circumstances. For example, is there flexibility on the final diagnostic after an AI software has made a suggestion if the clinician receives more information after the fact? Or does an AI prescription override doubt in the light of new information? \citet{emanuel2019artificial} argue that the promise of AI in healthcare is deliverable only if we are able to establish a positive behavioral approach in which patients trust the way their clinicians address the use of AI in their care. For example, healthcare providers can develop strategies that encourage empathy and trust in the digital realm \cite{terry2016emerging}, such as identifying and designing a set of fundamental ethical guiding principles \cite{taddeo2018ai, terry2019regulating} and utilizing AI in a way such that not only doctor-AI-patient interactions become more efficient, but doctor-patient interactions improve as well \cite{shirado2017locally, aminololama2019doctor}. Recent work has examined the potential of patient-centered approaches in AI-based healthcare \cite{findley2020keeping, bjerring2021artificial, ontika2022exploring}, showing that clinicians have a significant role in facilitating trust to normalize the use of AI systems and that the current lack of interpretability of the use of AI within healthcare hampers patient-centered medicine.

\subsection{Trust in AI}
Trust is generally understood as an essential element for the well-functioning of healthcare systems and clinician-patient relationships. Low trust levels are associated with a host of negative health and healthcare outcomes such as poorer health status, decreasing adherence to medications, and shorter relationships with doctors, among others \cite{graham2015influence}. Conversely, higher levels of trust are associated with better healthcare utilization and health outcomes, such as increased commitment to appointments and outpatient visits and decreased emergency visits \cite{whetten2006exploring, bodenlos2007attitudes}. 

Canalizing trust in AI-enabled healthcare has great potential for clinicians and patients alike. In terms of operational efficiency, it can decrease wait time for patients \cite{kennedy_2018} and decrease health-related information asymmetries \cite{sillence2004trust, weaver2009healthcare, ergonomics_2014, murray2003impact}. The latter is a step in the right direction towards increasing patient adherence to the prescribed treatment, in combination with a healthy relationship with the prescribing physician \cite{weaver2009healthcare, emanuel2019artificial, murray2003impact}. The use of AI in clinical support systems has clear advantages in diagnosis, treatment selection, and monitoring \cite{ngiam2019big,shaheen2021ai, yeasmin2019benefits}. There is also a positive consequence of the usage of support systems that offer relative anonymity to users and, along with other means of digital communication, reduce the reliance on face-to-face interaction between patients and carers. Anonymity has been proven to increase users’ self-disclosure on otherwise sensitive topics, including emotions and concerns \cite{clark2019anonymity}. This secondary feature is especially relevant for settings where health-related topics might be taboo and, therefore, difficult to discuss with carers for fear or shame. For instance, in India and Bangladesh, sexual and reproductive health is very much a taboo, and shame and stigma around it contribute to young girls’ negative health outcomes \cite{tellier2018menstrual}. Preliminary studies on the usage of health digital applications (e.g. mHealth \cite{labrique2018digital}) show that these tools are widely accepted by users in Bangladesh and India; however, digital literacy is still low, and in-person consultation still plays a crucial role in the development of better health service provision that is in accordance with local norms and patients’ ethics and values \cite{alam2017menstrual, messinger2017utilization}. Clinician-technology cooperation and interaction are key in these settings and can potentially impact how patients accept these technologies. 

\subsection{Technology Acceptance Models}
The theory behind the Technology Acceptance Model (TAM) was introduced by \citet{davis1989user} and outlines how users perceive and eventually come to accept technology. This model proposes that users rely on two factors when evaluating new technologies: the perceived usefulness (PU) of the technology and the perceived ease-of-use (PEOU) of the technology. Additionally, the authors find that external variables such as social influence have an impact on users' attitudes concerning a respective technology. %\juba{this seems like this is a causal model? good to describe that this is a causal approach here}

\begin{figure}[h]
\caption{The original Technology Acceptance Model (TAM) as outlined by \citet{davis1989user}.}
\centering
\includegraphics[width=0.35\textwidth]{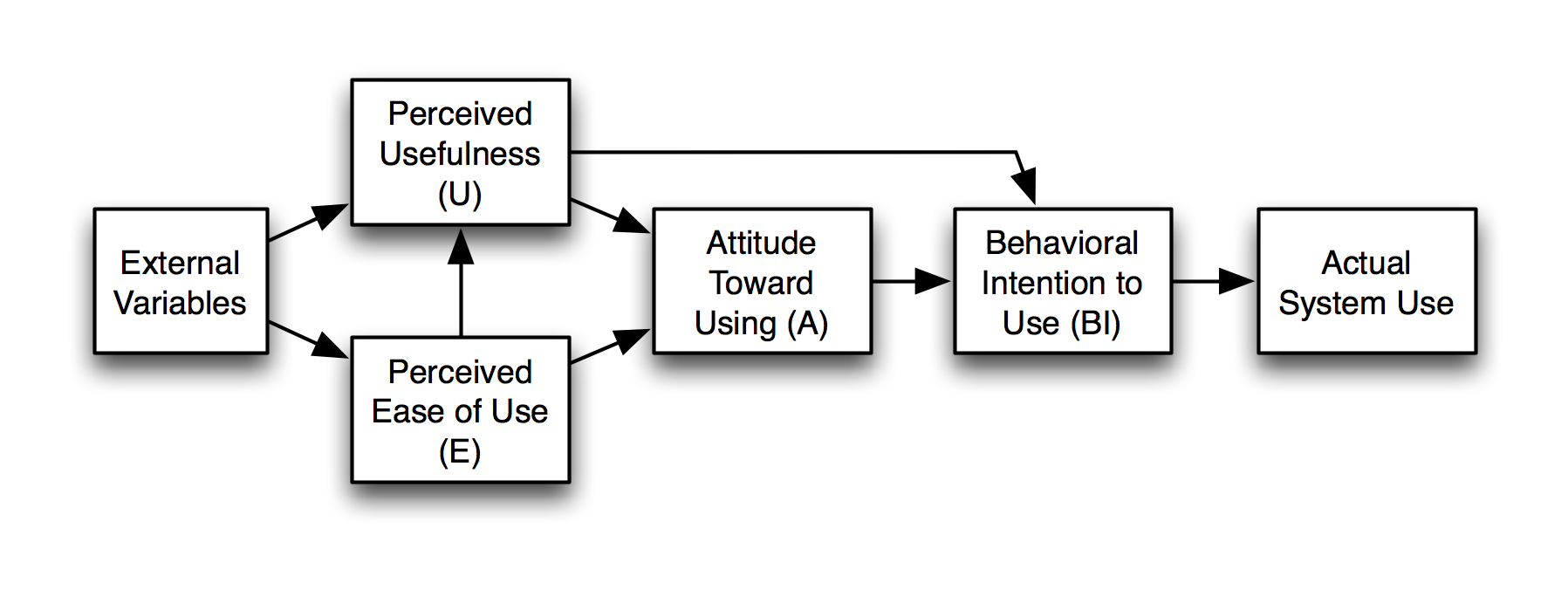}
\end{figure}

While TAM has been expanded to account for various social and cultural factors that influence technology acceptance, it continues to serve as a guide for technology integration into fields ranging from agriculture to education \cite{scherer2019technology}. Work by \citet{hu1999examining}, perhaps some of the earliest literature on technology acceptance for healthcare, expands on the TAM framework to examine physician reasoning behind their intentions to use telemedicine technologies. This work finds that, as a whole, TAM reasonably captures physicians' acceptance of new technologies, but some factors, such as perceived ease-of-use did not determine attitudes. This finding, along with work done by \citet{emad2016modified}, shows the need for a technology acceptance model able to capture the unique subtleties associated with technology acceptance within healthcare. \citet{emad2016modified} propose a modified technology acceptance model tailored towards healthcare informatics. They run quantitative experiments in the UK and Iraq using a model that includes traditional indicators, such as perception of the quality of the product, but also social variables, such as local norms. %\juba{quantitive experiments trying to achieve what? needs more description}. 
Again, perceived usefulness is found to be a significant factor in indicating the likelihood of technology acceptance. However, this work also highlights the need to examine cultural influences impacting the acceptance of novel technologies \cite{emad2016modified}. \citet{nadri2018factors} use the extended Technology Acceptance Model (TAM2) by \citet{venkatesh2000theoretical} in studies in Iran, further highlighting the need for TAMs to be tailored towards the respective environment and target populations of a healthcare setting where new technologies will be introduced. Continuing with the trend of TAM in healthcare, \citet{alhashmi2019implementing} further extend TAM for a study in the United Arab Emirates and find that factors such as the managerial, operational, and organizational makeup of healthcare facilities, along with their IT infrastructure impacts the integration of technology in this domain. \citet{dhaggara2020impact} shift from approaches taken by similar work and focus on the acceptance of technology in healthcare by patients rather than physicians. Their work augments TAM and shows that privacy and trust are also significant factors that impact technology acceptance, providing a solid base for the IAC framework introduced in this paper. As TAM continues to expand to different domains, more recent work has focused on the acceptance of AI as a technology. In their work, \citet{sohn2020technology} test a variety of technology acceptance theories, finding that the acceptance of AI-based technologies are more easily modeled by VAM, a values-based technology acceptance model incorporating factors such as enjoyment, usefulness, and economic feasibility \cite{kim2007value}, rather than the popular TAM framework \cite{sohn2020technology}. Our work contributes to prior literature on technology acceptance and expands on its applications to both healthcare and AI, which have not been explored as often.

\begin{figure}[h]
\caption{The Value-based Adoption Model by \citet{kim2007value}.}
\centering
\includegraphics[width=0.35\textwidth]{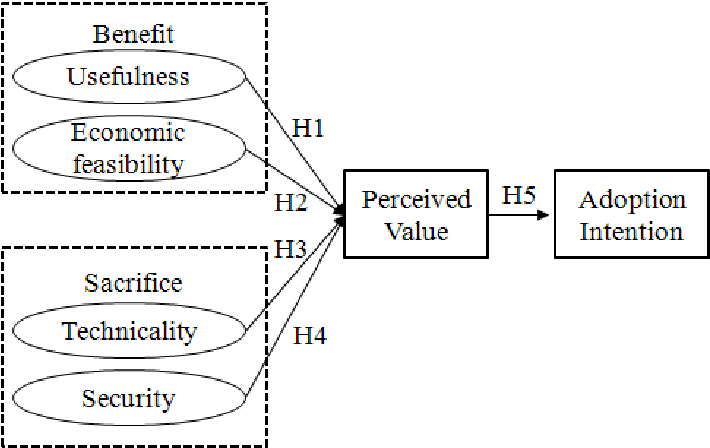}
\end{figure}

\subsection{Explainable AI}
Within the field of AI, explainability has become a major topic of interest. With the increasing complexity of the models used to train AI, it has become a struggle to comprehend decisions made by AI from the side of practitioners and end users \cite{hong2020human, caruana2020intelligible}. Within the field of explainable AI, a growing body of work is focused on examining the decisions made by AI and improving methods of making AI explainable to both technical experts and novice users. Work by \citet{hoffman2018metrics} addresses the question of the effectiveness in explaining AI systems to users and proposes measures that gauge the quality of explanations, user satisfaction with the explanations, user comprehension, and other factors such as the suitability of user trust and reliance on AI systems.
\citet{voosen2017ai}  highlights the challenge of embedding explainability into AI systems, demonstrating the tradeoffs that often occur between accuracy and ensuring that AI systems are transparent. XAI, a system developed by DARPA\footnote{The Defense Advanced Research Projects Agency (DARPA) is a research and development organization within the United States Department of Defense.} consists of three approaches to assist in the development of explainable AI models that maintain accuracy as the level of explainability increases \cite{gunning2017explainable}. Additionally, this system will ensure that human users both understand and trust AI. As the need for explainable AI heightens, a growing body of work has begun to focus on explainable AI for healthcare. \citet{lamy2019explainable} extend methods for analogical reasoning to propose visual techniques for classifying similar breast cancer cases that include both quantitative and qualitative similarities. Other work in the healthcare domain has explored the development of explainable AI tools for simulation training in surgery \cite{mirchi2020virtual}, predicting acute critical illnesses from electronic health records \cite{lauritsen2020explainable}, predicting low blood oxygen levels during surgery \cite{lundberg2018explainable}, digital pathology \cite{tosun2020explainable, jaume2020towards, uehara2020multi}, and much more. Motivated by the legal and privacy aspects of deploying AI within healthcare, \citet{holzinger2017we} propose methods for explainable AI, including expanding the capability of machine learning techniques to learn models that contain interpretable and causal features, introducing participatory methods where AI decisions would not overrule human-based medical decisions, and developing constructive user interfaces.

The explainability of AI should also consider the incentives and targets the particular algorithm is trying to optimize. In the context of image recognition, it is clear that the tools' outcome is (or, at the most, should be) to precisely identify a particular pattern of pixels linked to a medical condition. In the case of developing a diagnostic, however, it is possible for an AI application to include cost-cutting mechanisms in its target function. This means that a possible diagnosis or recommendation might push a patient in a specific direction without further consideration by a human, limiting the ability of the patient to make a fully informed decision. %\juba{last sentence is vague/not sure this is driving the point of how this can be a negative impact}

\citet{Becker_2019}, expanded by \citet{Bohr_Memarzadeh_2020}, suggests the four following goals for any healthcare-related artificial intelligence tool:
\begin{enumerate}
    \item Assessment of disease onset and treatment success.
    \item Management or alleviation of complications.
    \item Patient-care assistance during a treatment or procedure.
    \item Research aimed at the discovery or treatment of disease.
\end{enumerate}
Indicating that AI tools for healthcare aim to improve a patient’s life, \citet{Kakarmath_Esteva_Arnaout_Harvey_Kumar_Muse_Dong_Wedlund_Kvedar_2020} propose a list of best practices for research using AI. Our proposed framework aligns with these authors, incorporating techniques to make AI deployed within healthcare contexts "explainable" and expands work on explainable AI in healthcare, with a focus on improving both healthcare provider and patient understanding of AI.

\section{AI Typology}
% Don't forget to put table in here

\begin{table*}[h!]
\caption{AI Typology}
\begin{center}
\begin{adjustbox}{max width=\textwidth}
\begin{tabular}{lllll}
\cline{1-3}
\ & \textbf{Administrative} & \textbf{Diagnostic} &  &  \\ 
\cline{1-3}
\multicolumn{1}{l}{\textbf{Patient-facing}} &
  \multicolumn{1}{l}{\begin{tabular}[c]{@{}l@{}}
   Virtual Assistants for waiting rooms\\ 
   Health monitoring\\ 
   Medication management\\ 
   AI-driven robots assisting with care\end{tabular}} &
  \multicolumn{1}{l}{\begin{tabular}[c]{@{}l@{}}
   Computer-aided diagnosis (CADx, image processing)\\
   Digital consultations (clinical decision support systems, CDS)\\
   Brain-Computer interfaces (BCI)\\ 
   Group-level disease prevention\\
   Individualized precision medicine (with AI)\\
   Robot-assisted surgery\end{tabular}} &
   &
   \\ \cline{1-3}
\multicolumn{1}{l}{\textbf{Non patient-facing}} &
  \multicolumn{1}{l}{\begin{tabular}[c]{@{}l@{}}
   Bed allocations\\
   Scanning of Electronic Patient Records\\
   Staffing optimization: level and hiring\\
   Simulations for Medical Education\\
   Forecasting demand for medical resources\\
   Fraud detection\\
   Cybersecurity for health data\end{tabular}} &
  \multicolumn{1}{l}{\begin{tabular}[c]{@{}l@{}}
   Medical research and drug discovery (e.g. electrophysiology)\\ 
   Analyzing clinical laboratory results\end{tabular}} &
   &
   \\ \cline{1-3}
 &                         &                     &  &  \\
 \multicolumn{5}{l}{Note: authors' elaboration based on \citet{jiang2017artificial, arora2020conceptualising, kalis201810}. }
 \label{tab:AItypology}
\end{tabular}
\end{adjustbox}
\end{center}
\end{table*}

In order to achieve increased transparency in our approach to an explainable ethical framework, we introduce a patient-centered classification of Artificial Intelligence use in healthcare. Research on patients' attitudes towards AI in healthcare has highlighted three common elements across surveys:
\begin{enumerate}
    \item Pre-existing beliefs about a (healthcare) technology exist and affect its reception \cite{richardson2022framework}.
    \item Patients seem concerned about their ability to choose or communicate with their physician if AI is involved \cite{musbahi2021public,richardson2021patient}.
    \item Concerns about AI arise when there is little knowledge of its usage but are mitigated with improved knowledge communication \cite{musbahi2021public, richardson2021patient, richardson2022framework, scott2021exploring}.
\end{enumerate}

These items call attention to the crucial role communication and social interaction have in AI technology acceptance. To improve communication, we must first understand how and where AI is used in healthcare, i.e. refer to a healthcare AI typology. 

The more common AI typology in healthcare follows a classification based on input data type (structured and unstructured) and subsequent analytical method or departmental use of AI (e.g. workflow, care). However, when it comes to implementing an explainable ethical framework and behavioral protocol around AI social structures, we need to understand when AI comes in direct contact with patients or when it affects them indirectly through its role as an enabling technology. \Cref{tab:AItypology} presents a typology for AI in healthcare based on the amount of direct contact with patients. To the best of our knowledge, this is the first attempt at developing an AI typology that is patient-centered. Without a first attempt at recognizing the different ways and orders in which patients come in contact with AI in healthcare, we cannot begin to respond to the attitudinal issues raised above.
%\juba{ok so to play the devil's advocate here: there could be a million different ways to isolate elements that are relevant to ethics/behavior around AI social structures, including but not limited to the current one. What makes this compelling? Do we have evidence (e.g. prior work or studies) that those are first-order concerns that we need to pay attention to?}

Moreover, the introduction of such a typology system will allow researchers to focus on separate AI-enabled tools based on their functions in the healthcare context and how it changes physician-AI-patient interaction. It is a stepping stone towards the implementation of a transparent (or explainable) behavioral and ethical protocol, namely IAC. Following an interaction-based typology lets us categorize AI systems in use across hospitals, regions, and countries and, importantly, draw parallels between successful and non-successful deployments within the context of AI-human-patient interaction. This type of identification is especially relevant when patients believe their ability to interact with their caregivers is harmed \cite{musbahi2021public, scott2021exploring}, and there is still a general preference for human interaction \cite{young2021patient}.

% \begin{table}
% \caption{AI Typology}
% \label{table:AItype}
% \begin{tabularx}{\columnwidth}{ |X|X|X| } 
%  \hline
% \textbf{  } & \textbf{ Administrative }  & \textbf{ Diagnostic }\\
%  \hline
%  \textbf{(Patient-facing)} & \begin{itemize}
%             \item Health monitoring
%             \item Medication management
%             \item Virtual Assistants for waiting rooms
%             \item AI-driven robots assisting with care
%             \end{itemize}\\
%  \hline
%  \textbf{Non patient-facing} & \begin{itemize}
%             \item Computer-aided diagnosis (CADx, image processing)
%             \item Digital consultations (clinical decision support systems, CDS)
%             \item Brain-Computer interfaces (BCI)
%             \item AI-driven robots assisting with care
%             \item Group-level disease prevention
%             \item Individualized precision medicine (with AI)
%             \item Robot-assisted surgery
%             \end{itemize}
%  \hline
% \hline
% \end{tabularx}
% \end{table}

% Table + Caption -- Sources: Authors’ elaboration based on Artificial intelligence in healthcare: past, present and future (Jiang et al. 2018), Conceptualising Artificial Intelligence as a Digital Healthcare Innovation: An introductory Review (Arora, 2020), and Harvard Business Review’s 10 Promising AI Applications in Health Care (Kalis, Collier, Fu, 2018).
\section{IAC Framework}

\begin{figure*}[h!]
\caption{An outline of the three principles that construct the IAC framework and their respective implementation steps.}
\label{fig:iac_diagram}
\centering
\includegraphics[width=\textwidth]{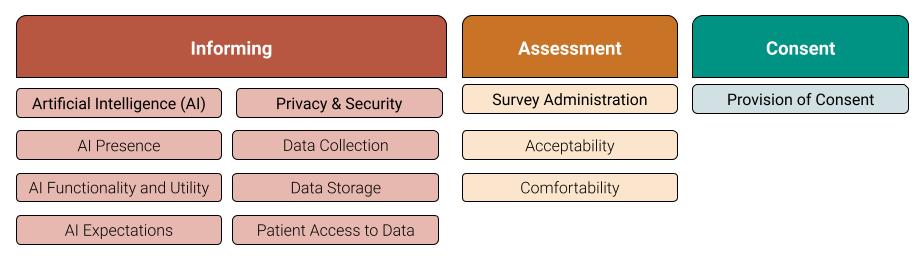}
\end{figure*}

\citet{larosa2018impacts} acknowledge the need for the establishment of standards that assist in maintaining trust between patients and healthcare providers. An extensive search of prior literature reveals a lack of standardization regarding the introduction of AI in healthcare \cite{reddy2020governance}. For AI-enabled technologies to have a significant impact on healthcare, we believe it is necessary for these technologies to be communicated to, reasonably understood by, and accepted by patients. In this paper, we propose a framework to guide how patients are informed about the presence of AI in their healthcare, measure patient acceptance and comfortability surrounding AI-enabled technologies, and assist in setting guidelines for how AI should be formally introduced to patients in healthcare settings. The core principles of this framework are broken down into 3 parts: {\sl Informing, Assessment}, and {\sl Consent}. For brevity, the framework will be referred to as IAC. The steps to implement the principles within IAC are detailed below and are also summarized within Figure \ref{fig:iac_diagram}.  

\subsection{Informing}
\subsubsection{Artificial Intelligence}
The lack of interpretability in AI systems is an issue that plagues both end users and developers alike \cite{hong2020human, caruana2020intelligible}. In the case of healthcare, these issues become amplified due to the presence of patients and the unique position they hold, being both non-AI and non-medical experts. With the advent of approaches focused on improving AI explainability within healthcare \cite{ahmad2018interpretable, stiglic2020interpretability, adadi2020explainable}, there is much promise for physicians and other medical practitioners to help close this gap by refining how they inform patients about the presence of AI and the accompanying security and privacy implications of using such technologies in patient care. We consider this process to be a form of ``explanation" where the physician explains the important functions of AI and what privacy and security measures will be enacted. With this in mind, we introduce the principle of \textit{Informing} to reshape existing approaches to medical transparency and impact patient trust.

\textbf{AI Procedure}
The \textit{Informing} principle is the first within the IAC framework and consists of informing the patient about AI and its respective privacy and security implications. In the implementation of this principle, the healthcare practitioner will begin by ensuring that the patient has been informed about the presence of AI in the equipment and/or software being directly used in their care. Following this, the patient will then be briefed on the purpose of the AI technology, what roles AI it aims to serve in patient care, and the exact functions it will perform throughout this process. From there, the healthcare practitioner will proceed to inform the patient about privacy and security.

\subsubsection{Privacy and Security}
% \juba{as a privacy person, I would potentially discuss inference attacks on outputs of computations (e.g. recovering private information without ever accessing it, just from the info encoded in the models we use. But this is a non-trivial discussion, and I feel this may be vastly out of scope}
Issues of privacy and security in healthcare are of extreme concern \cite{pramanik2019security, awotunde2021privacy, meingast2006security}. Over the past few years, data breaches such as hacking, malware attacks, and phishing have affected healthcare systems around the world almost incessantly \cite{mcleod2018cyber, seh2020healthcare, hammouchi2019digging}. As the need for machine learning systems to be trained on extraordinarily large amounts of data increases and the opportunities for contributing personal health data to these systems grow, it is imperative that patients are aware of the implications associated with engaging in AI-enabled technologies. When designing the {\sl Privacy \& Security} aspects of \textit{Informing} in the IAC framework, it was important to establish that patients would know what and how personal data is collected from AI-enabled technologies, how this data is securely and privately kept, and the autonomy that can be exercised in the management of their personal data. What is most important is that patients understand how they maintain agency over their own data. To maintain trust, especially when interacting with emerging technologies like artificial intelligence that are relatively complex, it may also be important for patients to have ownership of their personal healthcare data, where they are allowed to control privileges associated with access, use, storage, and possible deletion. While we posit this specific type of ownership as being important to patients, there is work needed to understand the limitations of such an approach in certain healthcare settings.

\textbf{Privacy and Security Procedure}
To implement the aspects of {\sl Privacy \& Security}, there are a few steps that healthcare practitioners can complete to ensure their actions are in alignment with the IAC framework. In this stage, the healthcare practitioner should begin by informing patients of what data is being collected from them while the AI system is being used in their care. During this step, the patient should be encouraged to ask questions pertaining to the data collection and storage process. Following this, the healthcare practitioner should explain how this data is collected and will describe where this respective data will be stored (locally in hospital data storage centers, in the cloud, etc.). Finally, the healthcare practitioner should ensure that the patient understands what access they have to their own data, along with explaining how this data can and will be used outside of their direct care. With respect to the {\sl Consent} principle of the IAC framework, patients should have the option to consent to the use of their data by second or third parties. This option aims to be exclusive of the consent given for the use of AI, meaning that patients should have the right to object to their data being used outside of their direct care (e.g. to train updates to AI systems) while still receiving treatment from the AI-enabled technology.

\subsection{Assessment}
To assess how patients perceive the use of AI in their healthcare after being informed about AI and its privacy and security implications, healthcare practitioners should assess how patients accept AI and how comfortable they feel about its use in their care. Based on the needs of the patient and the capabilities of the facility, the patient can be surveyed in verbal or written form using scales such as Likert, to which we propose examples below.

\subsubsection{Acceptability}
The condition of acceptability within healthcare contexts can provide significant insight into the effectiveness of a specific intervention but has been defined ambiguously throughout healthcare literature \cite{sekhon2017acceptability}. In their work, Sekhon et al. develop a comprehensive theoretical framework for acceptability that encompasses the subjective evaluations made by patients who experience and healthcare practitioners who deliver a respective intervention. With respect to the healthcare domain, \citet{dyer2016acceptability} synthesizes the term acceptability into a definition that embraces the aspects of experiential healthcare treatment and social validity. With this prior work in mind, we craft the principle of {\sl Acceptability} in the IAC framework to elucidate how AI-enabled technologies conform to a patient’s ethical values and expectations. It has been shown that patients are more receptive to treatment recommendations, leading to improved clinical outcomes \cite{hommel2013telehealth}. If patients are given the right to explicitly accept that a healthcare intervention enabled with AI will be used in their care, this will lead to greater trust in these respective systems and possibly enhanced quality of care.

%\juba{I think to have more of a punch, this needs a careful discussion of what is new here compared to the cited previous work of Sekhon et al.}

\textbf{Acceptability Procedure}
To continue using the IAC framework, the {\sl Acceptability} principle will be evaluated to measure how accepting a patient is of a proposed AI-enabled intervention being introduced for use in their healthcare. Our proposed evaluation methodology includes questions where the possible range of answers is based on a modified Likert scale with five options: Acceptable, Slightly acceptable, Neutral, Slightly unacceptable, and Unacceptable. While a Likert scale is proposed in this context, healthcare practitioners may find other measurement scales \cite{hodge2003phrase} more relevant to them. A sample list of questions is depicted below:

\begin{enumerate}
  \item How acceptable is having your healthcare practitioner use AI on you?
  \item How acceptable is this technology being used in your treatment?
  \item How acceptable is this technology being used in place of the practitioner?
  \item How acceptable is this technology being used to support the work of the practitioner? 
  \item How acceptable is this technology taking data from you? %\juba{would this potentially be a better fit under "Privacy and Security"?}
  \item How acceptable is this technology to have access to your personal health data? %\juba{same q as above}
  %\juba{\item something about role of transparency/explanations in acceptability? Would the technology be more acceptable if it gave you detailed explanation of the recs it makes? Something in those lines?}
  %\juba{\item something about acceptability of recs made by training on large amounts of data from similar patient vs expert human doctor? Which one is believed more? What if can argue that the AI recommendation are more informed than doctor?} 
\end{enumerate}
The above questions are a combination of affect and cognition measures \cite{sekhon2017acceptability} that may prompt informal questions about the technology from the patient to the healthcare provider. These conversations serve to ease the patients' worries and increase trust and transparency. 

\subsubsection{Comfortability}
Another important aspect of the IAC framework is to consider how comfortable patients are with AI-enabled technologies being incorporated and used within their healthcare. While comfortability is often overlooked in the informed consent process, it is a key part of the patient experience \cite{wensley2017framework}. Pelvic examinations are particularly known to be challenging experiences for both physicians and female patients and can be eased by using a plastic speculum (an instrument used to widen an orifice for inspection) over a metal one \cite{kozakis2006plastic}. Unfortunately, when receiving gynecological exams, there are cases where the selection of a speculum is left to the preference of the provider with little regard to patient choice \cite{bates2011challenging}. With the current state of AI integration in healthcare, we find stark similarities in the lack of patient choice. We believe that if patients are given the opportunity to express their level of comfort in how AI-enabled technologies are used in their treatment, this could lead to a stronger understanding of how patients perceive the use of AI in their care and may positively impact patient healthcare outcomes.

\textbf{Comfortability Procedure}
The {\sl Comfortability} stage of the IAC framework provides a place where a patient can express their respective feelings of comfort regarding the AI-enabled healthcare intervention to the healthcare practitioner. In this process, the healthcare practitioner can administer a qualitative survey consisting of a modified Likert scale to gauge how comfortable the patient is with the AI intervention and its associated implications. Similar to the principle of {\sl Acceptability}, in this survey, the healthcare practitioner will ask the patient a set of questions with the possible range of answers being Comfortable, Slightly comfortable, Neutral, Slightly uncomfortable, and Uncomfortable. As with \textit{Acceptability}, this proposed scale can be adapted to fit the domain context and patient needs. A list of sample questions is below:

\begin{enumerate}
  \item How comfortable are you with this healthcare practitioner using AI on you?
  \item How comfortable are you with this technology being used in your treatment?
  \item How comfortable are you with this technology being used in place of the practitioner?
  \item How comfortable are you with this technology being used to support the work of the practitioner? %\juba{flip side of this question: what if the AI only does recommendations but in the end the final decision making goes to a human expert/so technically not replacing the physician, just an additional support tool -- seems not mentioned below and echoes the support vs decision making tool discussion above?}
  \item How comfortable are you with this technology taking data from you?
  \item How comfortable are you with this technology having access to your personal health data?
\end{enumerate}

\subsection{Consent}

% \juba{Should this section have 5+ concrete questions to echo the structure and concreteness of the previous two? It seems more tentative/hand-wavy than 4.1 and 4.2.}

The informed consent process in healthcare has traditionally been opaque due to the lack of medical knowledge the average patient has, but recent advancements to improve this process have included multimedia interventions such as videos and interactive computer software or written information like informational leaflets or pamphlets \cite{paton2018impact}. More novel approaches, such as eConsenting (the practice of using digital devices such as mobile phones or computers in the informed consent process), have come into prominence across multiple areas of healthcare, but standards of this practice have yet to be implemented \cite{lunt2019electronic}. However, the process of informed consent is further convoluted by the lack of transparency provided when attempting to understand how or why certain equipment is being used in patient care. When consenting to receive medical treatment, patients generally consent to the overall process of receiving a respective treatment and have little to no liberty in the selection of the type of medical equipment or brands of medical supplies involved in their care. While considerable effort is taken upon physicians and other medical personnel to apprise patients of the possible benefits and/or detriments of a respective procedure, there is little transparency about the tools used in these procedures. When introducing the principle of {\sl Consent} in the IAC framework, ensuring that patients are informed about the presence and utility of AI within software or equipment being used in their care and that they consent to these technologies was of high importance.

\subsubsection{Consent Procedure}
After first briefing patients on AI and its respective implications and administering surveys to measure acceptability and comfortability, the healthcare practitioner can begin the process of seeking consent from the patient to proceed with using the AI-enabled technology in their care. It is important for the healthcare practitioner to acknowledge that consent is a dynamic process and can be revoked at any time. If the patient chooses not to go forward with this respective AI-enabled intervention, their choice should be respected, and a non-AI equivalent should be made available for use. With respect to privacy and security, patients should also have the option to consent to the use of their data by second or third parties. This option aims to be exclusive of the consent given for the use of AI, meaning that patients should have the right to object to their data being used outside of their direct care (e.g. to train updates to AI systems) while still receiving treatment from the AI-enabled technology. Throughout this process, we stress that the healthcare practitioner is transparent with the patient at all times in regard to how they illustrate the respective capabilities and risks of the AI intervention. Based on the available resources of the medical facility and operating procedures for obtaining authorization for medical care, consent from the patient can be confirmed through either a digital/handwritten signature or through verbal confirmation. 

%\juba{I think this paragraph is missing something. It addresses the "consent" part of informed consent, but not really the "informed" part. It seems to me that the "informed" part is really crucial for consent to have any form of validity here, and we might want to develop further mechanisms by which this informed consent can happen. We note above that "While considerable effort is taken upon physicians and other medical personnel to apprise
%patients of the possible benefits and/or detriments of a respective procedure, there is little transparency about the tools used in these procedures. When introducing the principle of Informed Consent in the IAC framework, ensuring that patients are informed about the presence and utility of AI within software or equipment being used in their care and that they consent to these technologies was of high importance." Can we build up on this? Make it less "vague"/provide more concrete steps that can be taken in that direction? Identify specific mechanisms in which transparency is lacking and how to improve on those, potentially referring to previous work?}

\begin{table*}
\begin{center}
  \caption{ A breakdown of the principles within the IAC framework and questions guiding their implementation.}
  \label{tab:commands}
  \begin{tabular}{p{3cm}p{10cm}}
    \toprule
    Principle & Related Question(s)\\
    \midrule
    Informing & Does the patient understand how AI will be used in their care? Does the patient understand the associated privacy and security implications of AI being used in their healthcare? \\
    Assessment & Does the patient accept that these technologies will be used on them/incorporated into their treatment? Is the patient comfortable with these technologies being used in their treatment?\\
    Consent & Does the patient consent to having these AI-enabled technologies used on them/incorporated into their treatment?\\
  \bottomrule
\end{tabular}
\end{center}
\end{table*}

\section{Case Study}
To conceptualize how the IAC framework would be deployed and to prepare for the various contexts in which IAC could be implemented, we will present three hypothetical scenarios below:

\subsection{Scenario I}
Client A is a large research hospital in the United States that recently bought a \$1.5 million AI system to install into their CT scanners to automatically scan for brain tumors. The radiologists and technicians employed in this hospital are well-versed in AI techniques and welcome the use of AI within their respective diagnostic workflows. The patients who visit this hospital are from middle to high-income backgrounds and are generally college-educated. The patients themselves are not experts in AI, but have heard or read about it in the news. Some patients work in technical fields as software engineers, data architects, etc., and may have firsthand experience with artificial intelligence through their professions. Client A has decided to adopt the IAC framework in alignment with recently mandated government regulations outlining the use of AI in medical equipment.

\subsection{Scenario II}
Client B is a clinic in the Philippines that has been collaborating with researchers from an industry research lab in their country. The researchers have developed an AI system to automatically scan for eye diseases and want to pilot it in the clinic. The nurses, the healthcare workers in charge of administering the eye scans, are unfamiliar with AI and have never used it before asides from apps with recommendation systems like Facebook or YouTube. The patients who visit this clinic work low to medium-wage jobs and have experience using technological devices such as mobile phones or computers. On average, the patients have no prior knowledge of AI and have similar experiences with using AI in apps to the nurses. In this pilot, the researchers from the industry lab want to test the IAC framework to gauge the potential of their system as a viable diagnostic tool for the clinic.

\subsection{Scenario III}
Client C is a community health program in rural Kenya that has received funding from a non-governmental organization to deploy AI-enabled smartphone applications that can screen for tuberculosis in their mHealth program. The smartphones are operated by community health workers (CHWs) who have the equivalent of a high school education and have received training from the NGO to conduct simple medical procedures such as measuring patient vital signs (breathing rate, blood pressure, body temperature, etc.) and screening for infectious diseases. The CHWs are familiar with using mobile phones through working with them on a daily basis. However, their knowledge of AI is extremely low or non-existent. On average, the patients involved in this community health program have the equivalent of a grade school-level education, and many of them work as day laborers in the agricultural sector or stay at home to take care of their children. The patients in this community have little experience with technology and have, at most, been exposed to basic (feature) mobile phones. The patients have had no exposure to AI and are unaware of it as a potential tool for disease diagnosis. The community health program has received guidance from the NGO to integrate IAC into the care routines of the CHWs and has received training from facilitators at the organization to do so. 

% \subsection{Ethical Concerns}
% One ethical concern arises when discussing the implementation of the empirical study. We acknowledge the disproportionate effects that could occur between the treatment and control groups. Providing training on the IAC framework to the treatment group will most likely put the hospitals in this group at an (unfair) advantage. For the control group, we expect patients to exit the study with a less thorough understanding of the AI system compared to their counterparts in the treatment group. We also expect that introducing IAC without the workshop will not have any detrimental effects on the healthcare outcomes of the patients in these respective medical facilities. At the moment, this is the primary ethical concern that we expect to face throughout our evaluation. As we iterate through developing this framework, we will ensure that experts with both medical and ethics backgrounds are actively involved. During the evaluation stage, we will also ensure that if available, compliance officers are present during these respective trainings to monitor the progress of these experiments.

\section{Limitations}
While the goal of IAC is to facilitate the transition of AI into healthcare, we understand that not all healthcare institutions have the capacity or ability to integrate these systems. We recognize the potential for those institutions with larger operating budgets and higher skilled physicians to have higher levels of access to these technologies, thus widening the already large healthcare gap seen between countries in the Global North and South and even within urban and rural areas within “developed” countries. However, we hope that as AI becomes more commonly integrated into healthcare systems around the world, IAC can serve as a guiding framework and be adapted to fit the distinct needs of healthcare institutions. Additionally, while IAC is meant to be used for all kinds of AI systems, we find that the framework may have to be expanded for contexts where smaller devices such as mobile phones and IoT devices such as monitoring bands, smart speakers, and sensors are commonly used in healthcare. 

\subsubsection{Bias in AI Technologies for Healthcare} 
Over the past few years, as the field of AI ethics has developed, concerns about bias within AI systems used for healthcare have become more prominent \cite{obermeyer2019dissecting, evans2020diagnosing, vyas2020hidden, celi2022sources}. Algorithmic bias is an issue that is more likely to affect marginalized populations and compound on existing inequities within fields such as healthcare, education, housing, and policing \cite{lee2018detecting, benjamin2019assessing, garcia2016racist}. As AI systems continue to evolve, we believe that it is imperative to address these issues to ensure that the decisions and solutions provided by these systems are, in fact, fair and transparent. We believe auditing systems for AI-powered technologies \cite{raji2020closing} or frameworks for assessing ethical considerations in healthcare technologies \cite{char2020identifying, fiske2020embedded, overton2020evaluation, baeroe2020machine} could be a better solution for directly addressing concerns of algorithmic bias within these systems. While the IAC framework does not directly address the issue of bias within AI software that is used directly for healthcare, we believe integrating the principles and methods from existing approaches will provide much more robustness within our respective framework. This is an area of future interest but not within the scope of our work at the moment.

\subsection{Socio-cultural Values}
While IAC provides a novel interaction-based approach to the concepts of acceptability, explainability, and privacy in AI for healthcare contexts, these concepts may not translate well to non-Western contexts. In regions where healthcare services are scarce due to the low number of medical professionals and facilities, the prospect of receiving urgent care through an AI intervention may trump the need for doctors to inform patients about the implications of these technologies. Research has also shown that concepts such as algorithmic fairness and privacy differ amongst users in Global South and Western contexts \cite{sambasivan2021re, al2020we, reichel2020have, naveed2022ask, chiou2009cross}. With this in mind, attempting to inform patients about why concepts such as data privacy or security are relevant to them may also be another task within itself to realize the full potential of IAC. Additionally, with both of the authors living in Western countries (the United States and the Netherlands, respectively), our values around privacy have been shaped by ideals that differ from those held by people living within the Global South. Working to understand how cultural perceptions of privacy, especially within healthcare contexts, impact the implementation of certain principles within IAC is an imperative task and will help shape future iterations of this framework.

% \subsection{Measurement Methodologies}
% At the moment, there are no known scales to measure the levels of acceptance, comfortability, or knowledge one has of AI-enabled technologies and applications, which introduces a bit of challenge to our approach. However, this shortcoming presents an opportunity to initiate the development of such methodologies. The metrics used to measure incorporate commonly used psychometric scales such as the Likert scale, which may not be complex enough to comprehensively capture the level of nuance needed to evaluate patient perception of AI. As we continue to develop the IAC framework and refine the experimental approach for our pilot evaluation, we will work to ensure that a scale that meets our needs is developed with these concerns in mind.

\subsection{Regulatory Limitations}
While AI and data protection regulation have advanced rapidly over the past decade \cite{butcher2019state, png2022tensions, okolo2023responsible, kak2020global, schmitt2022mapping}, few countries have formally enacted regulations that guide the use of AI in healthcare settings \cite{morley2022governing, vokinger2021regulating}. With this in mind, an additional hurdle that may impact the integration of IAC is the lack of regulatory structures in many countries outlining the use of AI in healthcare and other contexts. With no prior guide that governs the use of AI, it is unclear how IAC would complement existing incentive structures for hospitals and clinics to leverage the principles of our framework. However, this lack of regulation may benefit the IAC framework by impacting future medical school curricula and facility-specific healthcare AI guidelines. We also find that IAC has the potential to bridge these policy gaps and could inform policy-making in this context.

\section{Conclusion}
This paper introduces the IAC framework, which aims to improve how patients are informed about and consent to the use of AI-enabled medical technologies within their care. Our work leverages prior research in three areas: the social-behavioral foundations of technology in healthcare, technology acceptance models, and work in explainable AI to inform our approach. The IAC framework is motivated by the crucial role carer-patient trust plays in patients' health outcomes, and aims to guide how healthcare practitioners can inform patients about the use of AI in their healthcare, how practitioners can assess patients' acceptability and comfortability with the use of AI, and how patient consent can be gained after this process. While the field of human-AI interaction in the domain of artificial intelligence for healthcare is advancing, there is little work focused on examining how patients play an active role in these technologies. In particular, exploring the need for AI to be used in a complementary manner, where both patients and healthcare practitioners play an active role in the development and use of these technologies, could be beneficial to limiting instances of bias and improving how these technologies serve the best interests of patients. Ensuring that AI does not exacerbate existing disparities within healthcare begins with instituting guidelines and standards to monitor and guide its use. Our approach aims to diminish concerns that may arise from the introduction of AI in healthcare, but future work that leverages human-centered methodologies will be needed to fully understand patient perception and acceptance of AI.

% Future Work
% Our proposed framework provides a strong foundation to expand literature regarding the study of public concerns with AI-based technologies and human-centered design of these systems. To continue our work, we plan on deploying
% the empirical study outlined in this paper to test our proposed framework and evaluate its effectiveness in real-life settings. The insights of physicians, nurses, and other medical workers will be invaluable as we progress in expanding our research and we plan on forging a range of collaborative partnerships in this domain. Additionally, we find that there is no set order in which IAC should be applied, but as we experiment with the framework and improve the specifications of each principle, we may provide users with stricter guidelines dictating the application order of the principles within this framework.

\begin{acks}
We appreciate Dr. Juba Ziani for proofreading the paper and adding valuable insight that re-shaped our approach. 
\end{acks}

\bibliographystyle{ACM-Reference-Format}
\bibliography{ref}

\appendix

\end{document}